\documentclass[]{aa}
\usepackage{graphicx}
\usepackage{color}

\usepackage{txfonts}
\usepackage{natbib}
\bibpunct{(}{)}{;}{a}{}{,}
\definecolor{orange}{rgb}{1,0.5,0}

\def\mg{\ion{Mg}{I}}
\def\fe{\ion{Fe}{I}} 
\def\kms{$\mbox{km s}^{-1}$}

\def\co{$^{12}$CO\,(2--0)}
\def\dco{$D_{CO}$}

\def\ctwo{C$_{2}$}
\def\ctwoF{C$_{2}$}
\def\ctwoI{$C_{2}$}

\def\si{\ion{Si}{I}}
\def\al{\ion{Al}{I}}
\def\paper1{Paper~I}

\def\ngc1754{NGC\,1754}
\def\ngc2005{NGC\,2005}
\def\ngc2019{NGC\,2019}
\def\ngc1806{NGC\,1806}
\def\ngc2162{NGC\,2162}
\def\ngc2173{NGC\,2173}

\begin{document}

 \title{{Integrated $J$- and $H$-band  spectra of globular clusters in the LMC: 
 implications for stellar population models and galaxy age dating}
  \thanks{Based on observation collected at the ESO Paranal La Silla
    Observatory, Chile, Prog. ID 078.B-0205}}

\titlerunning{Integrated $J$- and $H$-band spectra of globular
  clusters in the LMC}

 \author{Mariya Lyubenova
          \inst{1}
          \and
          Harald Kuntschner\inst{2}
	  \and
	  Marina Rejkuba
	  \inst{2}
	  \and
	  David R. Silva
	  \inst{3}
	  \and
	  Markus Kissler-Patig
	  \inst{2}
	  \and
	  Lowell E. Tacconi-Garman
	  \inst{2}
          }

\authorrunning{M.~Lyubenova et al.}

 \institute{Max Planck Institute for Astronomy, K\"onigstuhl 17, D-69117 Heidelberg, Germany, \email{lyubenova@mpia.de}
 \and European Southern Observatory, Karl-Schwarzschild-Str. 2, D-85748 Garching bei
 M\"unchen, Germany, 
 \and National Optical Astronomy Observatory, 950 North Cherry Ave., Tucson, AZ, 85719
 USA 
}

\abstract {The rest-frame near-IR spectra of intermediate age (1-2
  Gyr) stellar populations are dominated by carbon based absorption
  features offering a wealth of information. Yet, spectral libraries
  that include the near-IR wavelength range do not sample a
  sufficiently broad range of ages and metallicities to allow for
  accurate calibration of stellar population models and thus the
  interpretation of the observations.}
{In this paper we investigate the {\em integrated}\/ $J$- and $H$-band
  spectra of six intermediate age and old globular clusters in the
  Large Magellanic Cloud (LMC). }
{The observations for six clusters were obtained with the SINFONI
  integral field spectrograph at the ESO VLT Yepun telescope, covering
  the $J$ $(1.09 - 1.41\,\mu$m) and $H$-band $(1.43 - 1.86\,\mu$m)
  spectral range. The spectral resolution is 6.7\,\AA\/ in $J$ and
  6.6\,\AA\/ in $H$-band (FWHM). The observations were made in natural
  seeing, covering the central $24\arcsec \times 24\arcsec$ of each
  cluster and in addition sampling the brightest eight red giant
  branch (RGB) and asymptotic giant branch (AGB) star candidates
  within the clusters' tidal radii. Targeted clusters cover the ages
  of $\sim$1.3\,Gyr (NGC\,1806, NGC\,2162), 2\,Gyr (NGC\,2173) and $\sim$13\,Gyr
  (NGC\,1754, NGC\,2005, NGC\,2019). }
{$H$-band \ctwo\/ and $K$-band \co\/ feature strengths for the LMC globular clusters are compared to the models of Maraston (2005). \ctwo\/ is reasonably well reproduced by the models at all ages, while \co\/ shows good agreement for older (age $\ge$~2 Gyr) populations, but the younger (1.3 Gyr) globular clusters do not follow the models. We argue that this is due to the fact that the empirical calibration of the models relies on only a few Milky Way carbon star  spectra, which show different \co\/ index strengths than the LMC  stars. The \ctwo\/ absorption feature strength correlates strongly
  with age. It is present essentially only in populations that have
  1-2 Gyr old stars, while its value is consistent with zero for older
  populations. The distinct spectral energy distribution observed for
  the intermediate age globular clusters in the $J$- and $H$-bands
  agrees well with the model predictions of Maraston for the
  contribution from the thermally pulsing asymptotic giant branch
   phase.}
{In this pilot project we present an empirical library of six LMC
  globular cluster integrated near-IR spectra that are useful for
  testing stellar population models in this wavelength regime. We show
  that the $H$-band \ctwoF\/ absorption feature and the $J$-, $H$-band
  spectral shape can be used as an age indicator for intermediate age
  stellar populations in integrated spectra of star clusters and galaxies.
 }


\maketitle

\section{Introduction}
\label{sec:motivation}
With the appearance of more and better near-IR instruments, the
studies of the stellar populations in galaxies in different
environments and at various redshifts, exploring the rest-frame
wavelength region between 1 and 2.5~$\mu$m, have become more frequent
\citep[e.g.][]{silva2008, lancon08,hyvonen09,esther09,cesetti09,martins10,riffel11b}. Thus
a reliable framework to interpret these observations is needed. The
comparison of observations with stellar population models is an
approach that proved to be successful in the optical wavelength regime
over the last decades. However, the near-IR range of the evolutionary
populations synthesis (EPS) models still lacks a proper empirical
calibration.  Recent advances in theoretical as well as empirical calibration include in particular work on asymptotic giant branch (AGB) evolutionary phase, whose contribution is crucial to the near-IR part of the models, \citep[e.g.][]{marigo08,girardi10,maraston98,maraston05}. In \citet[][hereafter \paper1]{az10a} we described our efforts to create a near-IR library of integrated spectra of globular
clusters to serve as a calibrator and test bench for existing and future EPS models \citep[e.g.][]{bruz03,maraston05,con2009,maraston11}. For this pilot project we chose six globular clusters in the LMC, for which detailed information about the age and chemical composition exists, based on
their resolved light. Our sample consists of three old ($>$~10~Gyr)
and metal poor ([Fe/H]~$\simeq -1.4$) clusters, namely NGC\,1754,
NGC\,2005, and NGC\,2019, as well as three intermediate age $(1 -
2$\,Gyr) and more metal rich ([Fe/H]~$\simeq-0.4$) clusters:
NGC\,1806, NGC\,2162, and NGC\,2173 (see Table~\ref{tab:lmc_gc}).

%
%
\begin{table*}[tdp]
\caption{\label{tab:lmc_gc}Target globular clusters in the LMC.}
\centering
\begin{tabular}{c c c c c c c c c c c}
\hline 
\hline 
Name & Age (Gyr) & [Fe/H] & $L_{mosaic}$ ($L_{\sun}$) & $r_{h}$ & $r_{t}$ & S/N$_{J}$ & S/N$_{H}$\\ 
(1) & (2) & (3)  & (4) & (5) & (6) & (7) & (8)\\
\hline
NGC\,1806  & 1.1\tablefootmark{1}, 1.7\tablefootmark{8}, 1.5\tablefootmark{9}, 1.9\tablefootmark{2}, 0.5\tablefootmark{e}  & -0.23\tablefootmark{b}, -0.71\tablefootmark{e} & $3.4\times10^{4}$ & -- & -- & $>$60\tablefootmark{*} & $>$45\tablefootmark{*}\\ 
NGC\,2162  &  1.1\tablefootmark{1}, 1.3\tablefootmark{3}, 2\tablefootmark{2}, 0.9\tablefootmark{4}, 1.25\tablefootmark{5}  & -0.23\tablefootmark{b}, -0.46\tablefootmark{f} & $0.5\times10^{4}$ & 21\farcs37 & 197\farcs2 & $>$45\tablefootmark{*} & $>$40\tablefootmark{*}\\ 
NGC\,2173  & 2\tablefootmark{1}, 2.1\tablefootmark{3}, 1.6\tablefootmark{6}, 4.1\tablefootmark{2}, 1.5\tablefootmark{7}, 1.1\tablefootmark{4}, 1.6\tablefootmark{5}  & -0.24\tablefootmark{b}, -0.42\tablefootmark{f}, -0.51\tablefootmark{g} & $0.9\times10^{4}$ & 34\farcs35 & 393\farcs5 & $>$80\tablefootmark{*} & $>$45\tablefootmark{*}\\
NGC\,2019  & 10\tablefootmark{1}, 16\tablefootmark{10}, 13.3\tablefootmark{11}, 16.3\tablefootmark{a}, 17.8\tablefootmark{b} & -1.23\tablefootmark{a}, -1.18\tablefootmark{b}, -1.37\tablefootmark{c}, -1.10\tablefootmark{d} & $6.7\times10^{4}$ & 9\farcs72 & 121\farcs6 & 95 & 150\\
NGC\,2005  & 10\tablefootmark{1}, 6.3\tablefootmark{10}, 16\tablefootmark{11}, 15.5\tablefootmark{a}, 16.6\tablefootmark{b} & -1.35\tablefootmark{a}, -1.92\tablefootmark{b}, -1.80\tablefootmark{c}, -1.33\tablefootmark{d} & $3.9\times10^{4}$ & 8\farcs65 & 98\farcs8 & 140 & -- \\
NGC\,1754  & 10\tablefootmark{1}, 7\tablefootmark{10}, 14\tablefootmark{11}, 15.6\tablefootmark{a,b} & -1.42\tablefootmark{a}, -1.54\tablefootmark{b} & $2.7\times10^{4}$ &  11\farcs2 & 142\farcs9 & 80 & --\\
\hline
\hline
\end{tabular}
\tablefoot{
(1) Cluster name, (2) age of the cluster in Gyr, derived using different methods:
\tablefoottext{1} {\citet{frogel90} -- based on the SWB type;}
\tablefoottext{2} {\citet{leonardi03} -- integrated spectroscopy;}
\tablefoottext{3} {\citet{geisler97},}
\tablefoottext{4} {\citet{girardi95},}
\tablefoottext{5} {\citet{kerber07},}
\tablefoottext{6}{\citet{bertelli03},}
\tablefoottext{7}{\citet{woo03},}
\tablefoottext{8}{\citet{mackey08},}
\tablefoottext{9}{\citet{milone09} -- all CMDs,}
\tablefoottext{10}{\citet{beasley02} -- integrated spectroscopy (H$\beta$-Mg$b$),}
\tablefoottext{11}{\citet{beasley02} -- integrated spectroscopy (H$\gamma$-$\langle$Fe$\rangle$),}
(3) [Fe/H] derived using different methods:
\tablefoottext{a}{\citet{olsen98} -- slope of the RGB,}
\tablefoottext{b}{\citet{olszewski91} -- low-resolution Ca\,II triplet,}
\tablefoottext{c}{\citet{johnson06} -- high-resolution Fe\,I,}
\tablefoottext{d}{\citet{johnson06} -- high-resolution Fe\,II,}
\tablefoottext{e}{\citet{dirsch00} -- Str\"{o}mgren photometry,}
\tablefoottext{f}{\citet{groch06} -- low-resolution Ca\,II triplet,}
\tablefoottext{g}{\citet{muc08} -- high-resolution spectroscopy}
(4) sampled bolometric luminosity within the clusters central mosaics in $L_{\sun}$, (5) half-light radius and (6) tidal radius of the King-model cluster fit from the catalogue of \citet{mclaughlin05}, (7) 
and (8) S/N of the integrated spectra in $J$- and $H$-band, respectively;
\tablefoottext{*}{lower threshold, see Sect.~\ref{sec:spec_int} for an explanation}.
}
\end{table*}

In \paper1\/, using the integrated $K$-band spectra of the globular
clusters, we demonstrated the feasibility of our observational
approach, as well as discussed some discrepancies that arise between
recent EPS models and the data. We argued that the main reason for
these discrepancies is the incompleteness of the current stellar
spectral libraries of asymptotic giant branch (AGB) stars that are
used to create the models. Also, we illustrated how the presence of luminous carbon-rich
AGB stars significantly changes the observed spectrophotometric
properties in the $K$-band, especially the \co\/ 2.29 $\mu$m feature as predicted by \citet{maraston05}, despite the observational discrepancies.

This paper complements \paper1\/ by presenting results based on the
integrated $J$- and $H$-band spectra of the same globular clusters.
To date there are very few studies of the spectral properties of
globular clusters in this wavelength regime
\citep[e.g.][]{riffel11a}. The paper is organised as follows: in
Sect.~\ref{sec:obs} we briefly recall our observing strategy and data
reduction methods. In Sect.~\ref{sec:sed} we compare the observed
overall spectral energy distributions of the globular clusters to
stellar population models. Sect.~\ref{sec:c2} is devoted to the
\ctwo\/ absorption feature in the $H$-band and its behaviour in the
spectra of globular clusters. In Sect.~\ref{sec:discussion} we discuss
potential reasons for the disagreement between model predictions and
our observations, based on stellar libraries. Finally, in
Sect.~\ref{sec:conclusions} we present our concluding remarks.

\section{Observations and data reduction}
\label{sec:obs}
\subsection{Observations}
Our sample of globular clusters, whose basic properties are listed in
Table~\ref{tab:lmc_gc}, were observed with VLT/SINFONI
\citep{eis03,bonnet04} in poor seeing conditions without adaptive
optics in the period October -- December 2006 (Prog. ID 078.B-0205,
PI: Kuntschner). The detailed descriptions of the observing strategy,
target selection and data reduction are given in \paper1\/, where we
presented the $K$-band spectra from our project. In the present paper
we explore the data obtained in the $J$- and $H$-bands, where minor
differences in the data reduction exist. Briefly, in order to sample
at least the half-light radius of the clusters, we observed a
$3\times3$ mosaic of the largest field-of-view that SINFONI offers
($8\arcsec \times 8\arcsec$). Thus, the resulting coverage was
$24\arcsec \times 24\arcsec$, with a spatial sampling of 0\farcs125
per pixel. For the observations we used the standard near-IR nodding
technique. We observed each mosaic tile three times with individual
integration times of 50\,s and dithering of 0\farcs25 to ensure better
bad pixel rejection. We took sky exposures between each mosaic
tile. Finally, within each cluster's tidal radius, we selected up to
eight bright stars with colours and magnitudes typical for the red
giant branch (RGB) and AGB stars in the LMC (see Table 3 in \paper1),
which were observed in addition to the cluster mosaics in order to
better sample the short lived AGB phase. The integration times for
these stars were 10\,s. After each science target and at a similar
airmass, we observed an A-B dwarf star for telluric absorption
correction (Table 4 in \paper1).

%
%
\begin{figure*}
\resizebox{\hsize}{!}{\includegraphics[angle=0]{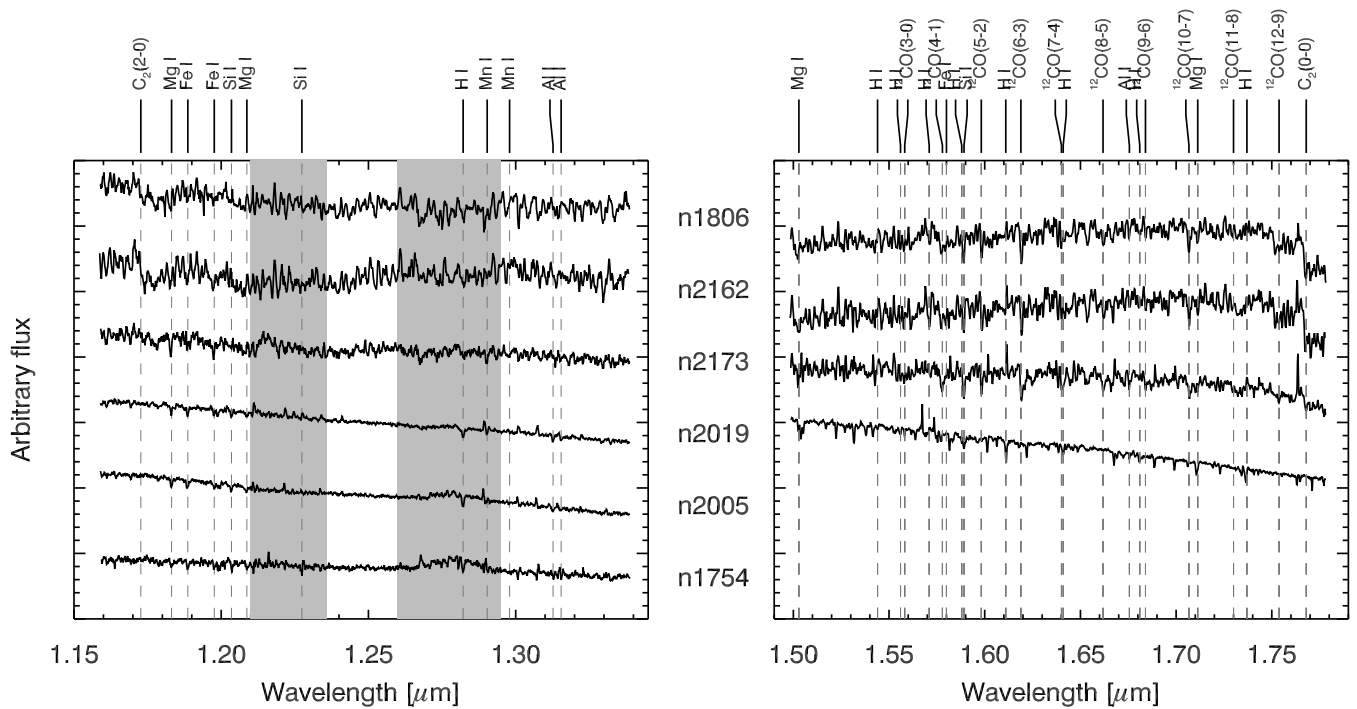}}
\caption{\label{fig:jh_spec} {\it Left panel:} $J$-band, integrated
  spectra of the six LMC globular clusters. The shaded areas indicate
  the regions with contamination by strong OH and O$_{2}$ sky line
  residuals. {\it Right panel:} $H$-band, integrated spectra of four
  of the LMC clusters. Due to unstable atmospheric conditions, the
  $H$-band spectra of NGC\,1754 and NGC\,2005 are heavily contaminated
  by sky line residuals and thus exibit a very low S/N ratio and are
  not shown. Each spectrum is normalised to its median value. Line
  identifications are based on the stellar spectral atlases by
  \citet{lancon92},\citet{wallace00} and \citet{rayner09}. }
\end{figure*}
\subsection{Data reduction}
\label{sec:data_red}
We used the SINFONI pipeline version 2.0.5 to perform the basic data
reduction on the three exposures per mosaic tile plus two bracketing
sky exposures. In brief, the pipeline extracts the raw data, applies
distortion, bad pixel and flat-field corrections, wavelength
calibration, and stores the combined sky-subtracted spectra in a
3-dimensional data cube. Then on each resulting data cube we ran the
{\tt lac3d} code \citep{davies10}, whose purpose is to detect and
correct residual bad pixels that are identified using a 3D Laplacian
edge detection method. This code not only removes residual bad pixels,
but also produces an error data cube for each input science data cube,
which is very useful, because the SINFONI pipeline does not provide an
estimate of the error propagation during data reduction. The derived
signal-to-noise ratio, using these error spectra, is in agreement with
the signal-to-noise ratio derived using an empirical method described
by \citet{stoehr07}.

In the $J$-band we had to switch off the pipeline option to correct
for sky line residuals based on the algorithm described in
\citet[][i.e. {\tt --objnod-sky\_cor = FALSE}]{davies07}. This was
needed due to the complicated pattern of overlapping OH and O$_{2}$
sky emission lines in this wavelength region \citep[see
e.g.][]{rousselot00}, which leads to their over- or under-subtraction
when applying the sky residual correction. Therefore, we applied
simple sky subtraction and indicated the regions where the sky lines
are the strongest in Fig.~\ref{fig:jh_spec} (i.e. 1.21\,--\,1.236
$\mu$m and 1.26\,--\,1.295 $\mu$m).

We reduced the telluric star data in the same way as the science
frames. Then for each telluric star we extracted a one-dimensional
spectrum, removed the hydrogen absorption lines fitted with a
Lorentzian profile, and divided the star spectrum by a black body
spectrum with the same temperature as the star. The last step in
preparing the telluric correction spectrum was to apply small shifts
($<$0.05 pixels) and scalings to minimise the residuals of the
telluric features. Finally, we divided each science data cube by the
corresponding telluric spectrum. In this way we also obtained an approximate
relative flux calibration. An absolute flux calibration was not possible due to non-photometric conditions.

The telluric star HD\,44533, used for the telluric correction of the
observations for NGC\,2019 and its surrounding stars, had an unusual
shape of the hydrogen lines. It seems that this star shows some
hydrogen emission together with absorption. Thus, to remove the
hydrogen lines, we interpolated linearly over the affected regions,
which are not overlapping with any indices used in this study.
However, we add a word of caution to treat these regions with care in
case the spectra of NGC\,2019 are used in the future for other
purposes.

\subsection{Spectra integration}
\label{sec:spec_int}
To obtain the final, integrated $J$- and $H$-band spectra for the six
globular clusters, we estimated the noise level in each reconstructed
image from the central mosaic data cubes. We assumed that this noise
is due to residuals after the sky background correction. Thus, we
computed the median residual sky noise level and its standard
deviation, after clipping all data points with intensities of more
than 3$\sigma$. We then selected all spaxels, which have an intensity
more than three times the standard deviation above the median residual
sky noise level. We summed them and normalised the resulting spectrum
to 1\,s of exposure time. The same approach was used to obtain the
spectra for the additional bright RGB and AGB stars.

The final step in preparing the luminosity weighted, integrated
spectra for the six globular clusters was to add the additionally
observed bright stars outside the central mosaics (for a detailed
discussion on cluster membership and which stars where included, see
Sect.~4 in \paper1\/ and Table~\ref{tab:bs_observations}). Fig.~\ref{fig:jh_spec} shows the
final, integrated $J$- and $H$-band spectra for the six LMC globular
clusters.  The spectra of the three intermediate age clusters are
shown on the top (NGC\,1806, NGC\,2162, and NGC\,2173).  These spectra
display numerous absorption features due to carbon based molecules,
like CO, \ctwo, and CN which makes them appear to be very noisy, while in
fact they are not. The spectra of the old and metal poor clusters
(NGC\,2019, NGC\,2005, and NGC\,1754) are shown at the bottom. These
are dominated by late-type giant stars, where the metallic lines are
more prominent. To identify the spectral absorption lines displayed in
Fig.~\ref{fig:jh_spec}, we used the stellar spectral atlases of
\citet{lancon92}, \citet{wallace00}, and \citet{rayner09}. The weather
conditions during the observations of the $H$-band spectra for
NGC\,1754 and NGC\,2005 were not favourable, which resulted in a
signal-to-noise ratio that was too low and contamination by multiple
sky line residuals. Thus we decided to exclude these two spectra from
our analysis.

The spectral resolution of our spectra, as measured from arc lamp
frames, is 6.7\,\AA\/ and 6.6\,\AA\/ (FWHM) in the $J$ and $H$-band,
respectively.  Finally, we estimated the S/N ratio of each integrated
globular cluster spectrum using the method of \citet{stoehr07} and
listed the values in Table~\ref{tab:lmc_gc}. This method allows us to
compute the S/N from the spectrum itself. However, due to the numerous
absorption features mimicking noise in the spectra of globular
clusters containing carbon-rich AGB stars, i.e. the three intermediate
age clusters marked with an asterisks in Table~\ref{tab:lmc_gc}, we
are able to give only a lower threshold. Based on the integration
times and luminosities of the clusters, we conclude that the quality
of these intermediate age spectra is as good, if not better, than
those of the old globular clusters.

\section{The overall near-IR spectral energy distribution}
\label{sec:sed}
%
%
%
%
\begin{figure*}
\resizebox{\hsize}{!}{\includegraphics[angle=0]{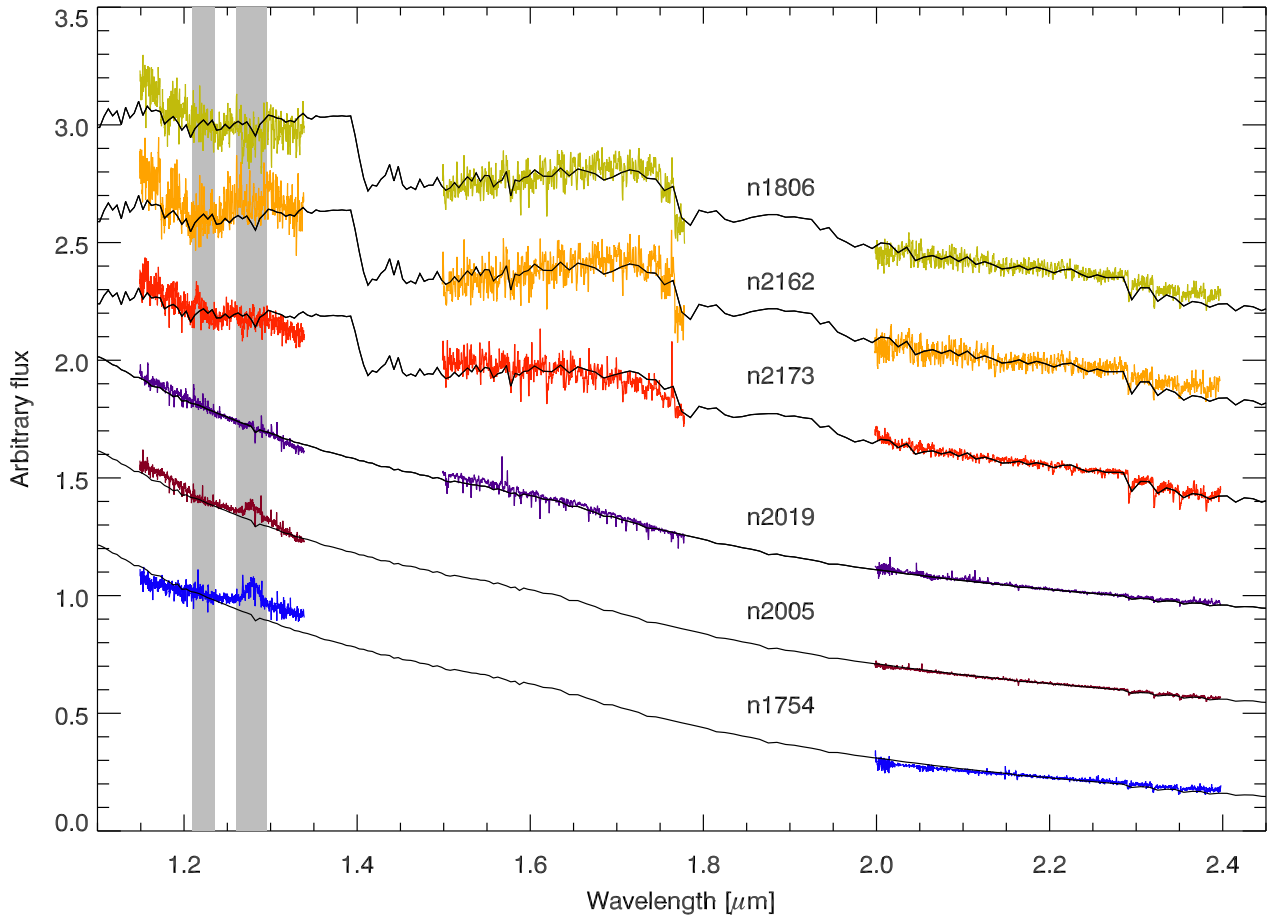}}
\caption{\label{fig:lmc_sed} $J$, $H$, $K$-band spectral energy
  distributions of our sample of six LMC globular clusters. The black
  solid lines indicate the most closely matching (in age and
  metallicity) stellar population model of \citet{maraston05};
  i.e. age~=~13\,Gyr, $[Z/H]=-1.35$ for NGC\,1754, NGC\,2005, and
  NGC\,2019, age~=~2\,Gyr, $[Z/H]=-0.33$ for NGC\,2173, and
  age~=~1.5\,Gyr, $[Z/H]=-0.33$ for NGC\,1806 and NGC\,2162. The spectra
  are normalised to the flux at 1.25~$\mu$m. Note the prominent
  spectral shape differences in the $J$- and $H$-bands between old and
  intermediate age clusters.  The shaded areas indicate the regions with contamination by strong OH and O$_{2}$ sky line residuals. }
\end{figure*}

Having completed the primary goal of our project, namely to provide an
empirical library of integrated near-IR spectra of globular clusters,
we show in Fig.~\ref{fig:lmc_sed} the overall $J$, $H$ and $K$-band
 spectra as derived from our SINFONI observations, compared to the spectral energy distributions from the \citet{maraston05} stellar population models. The observed spectral segments were scaled individually relative to the model providing the closest match in age and metallicity and were normalised to unity at 1.25~$\mu$m  (see Fig. caption).
In the figure the models are represented with solid, black lines.  With the
exception of the $J$-band part in old globular clusters, where we were
unable to achieve good sky background removal, as explained in
Sec.~\ref{sec:data_red}, the models agree well with the general
spectral shape of our spectra.  Remarkably, the distinct spectral
energy distribution (the "sawtooth" pattern) in the $J$- and $H$-bands
caused by the contribution from carbon-rich thermally pulsing asymptotic giant
branch (TP-AGB) phase is visible in the spectral regions covered by SINFONI for the intermediate age clusters. As expected the pattern is  less prominent for the cluster NGC\,2173 due to its slightly older age.  Also, the most prominent features such as \ctwo\/ at 1.77~$\mu$m and the CO bandhead at 2.29~$\mu$m are well matched.  Thus, when they become available, it will be very
interesting to make a comparison with higher spectral resolution EPS
models. Weaker features, as the ones indicated in
Fig.~\ref{fig:jh_spec}, could be studied in detail. In this figure, in
addition to carbon based molecular absorption bands, several other
absorption features are indicated.  They are mainly due to metallic
lines, such as \fe\/, \mg\/, \si\/, and \al\/. However, due to their
relative weakness and associated difficulty of measuring them in
galaxies, we decided not to discuss them further in this paper.

\section{The $H$-band \ctwoI\/ index}
\label{sec:c2}
One of the most prominent and easy to quantify features in the
$H$-band is the C$_{2}\,(0-0)$ bandhead at 1.77\,$\mu$m
\citep{ballik63}. The existence of the \ctwoF\/ molecule is typical for
carbon-rich stars, where the ratio of carbon to oxygen (C/O) atoms is
larger than 1. This type of star is the main contributor to the
near-IR light of intermediate age (1-3\,Gyr) stellar populations \citep[e.g.][]{ferraro95,girardi98,maraston98,maraston05}.
 Thus, it is expected that the \ctwoF\/ absorption
feature will be strong in globular clusters and galaxies with stellar
populations in this age interval \citep{lancon99,maraston05}.
 Also, more metal poor stellar
populations exhibit a stronger \ctwoI\/ index on average, because dredge-up more easily leads to C/O$>$1 in that regime \citep{wagenhuber98,mouhcine03,weiss09}.

\begin{figure}
\resizebox{\hsize}{!}{\includegraphics[angle=0]{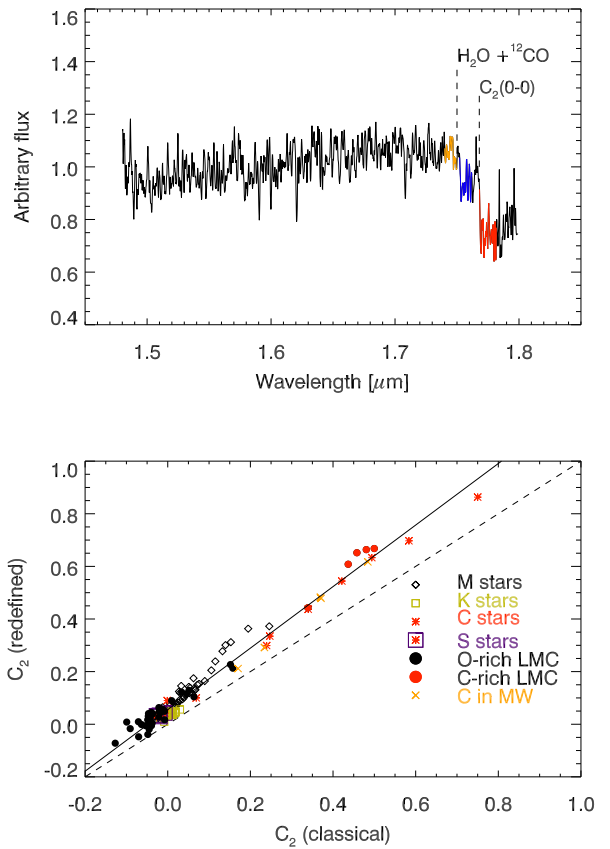}}
\caption{\label{fig:c2_c22_comp} {\it Top panel:} Spectrum of NGC\,1806 with
  overplotted central passband (red) and the classical continuum
  region (blue) for the \ctwoI\/ index \citep[][]{maraston05}. In
  orange we overplot the new continuum region for the \ctwoI\/ index
  defined in this paper (see Sect.~\ref{sec:c2}).  {\it Bottom panel:}
  Comparison between the classical and the redefined \ctwoI\/ index,
  measured on our own LMC star observations (solid black and red
  symbols), as well as the stars from the Milky Way spectral libraries
  of \citet[][orange]{lancon02} and \citet[][black diamonds, green
  squares, red asterisks]{rayner09}. The dashed line shows the
  one-to-one relation, the solid line represents a linear least
  squares fit to all data points.}
\end{figure}

The \ctwoI\/ index reflects the depth of the \ctwoF\/ absorption feature
at 1.77\,$\mu$m \citep{alvarez00}. In our work we used the definition
employed by \citet[][hereafter we refer to this definition as
``classical'']{maraston05}. The index is defined in magnitudes, as the
flux ratio between a central passband ($1.768 - 1.782\,\mu$m) and a
continuum passband ($1.752 - 1.762\,\mu$m), and is normalised to Vega
by subtracting a zero point of 0.038$^{m}$. In this definition of the
index, the continuum band is on top of H$_{2}$O and $^{12}$CO features
(see Fig.~\ref{fig:c2_c22_comp}, top panel). While carbon-rich stars are not expected to have  H$_{2}$O absorption, these two features may vary with the pulsation period for oxygen-rich stars \citep{loidl01}.

%
%

Because of these additional contributions to the continuum passband we
explored  a modified index definition by  shifting the continuum passband to shorter wavelengths relative
to the classical definition ($1.74 - 1.75\,\mu$m, shown in orange in
Fig.~\ref{fig:c2_c22_comp}). The Vega zero point to be subtracted is
0.037$^{m}$. We measured the two indices for the LMC stars from our
own observations, the carbon-rich averaged stellar spectra of
\citet{lancon02} and the stars from the library of
\citet{rayner09}. The comparison is shown in
Fig.~\ref{fig:c2_c22_comp} (bottom panel). We found that for these
samples the scatter is not significant, but one might expect it in a
larger sample, and in particular if one includes AGB stars in
different pulsation phases. For the present study we decided to use
the classical index definition for direct comparison to models. The
measurements for the stars in our library are listed in
Table~\ref{tab:bs_observations} and for the globular clusters in
Table~\ref{tab:lmc_c2}.

%
%
\begin{figure}
\resizebox{\hsize}{!}{\includegraphics[angle=0]{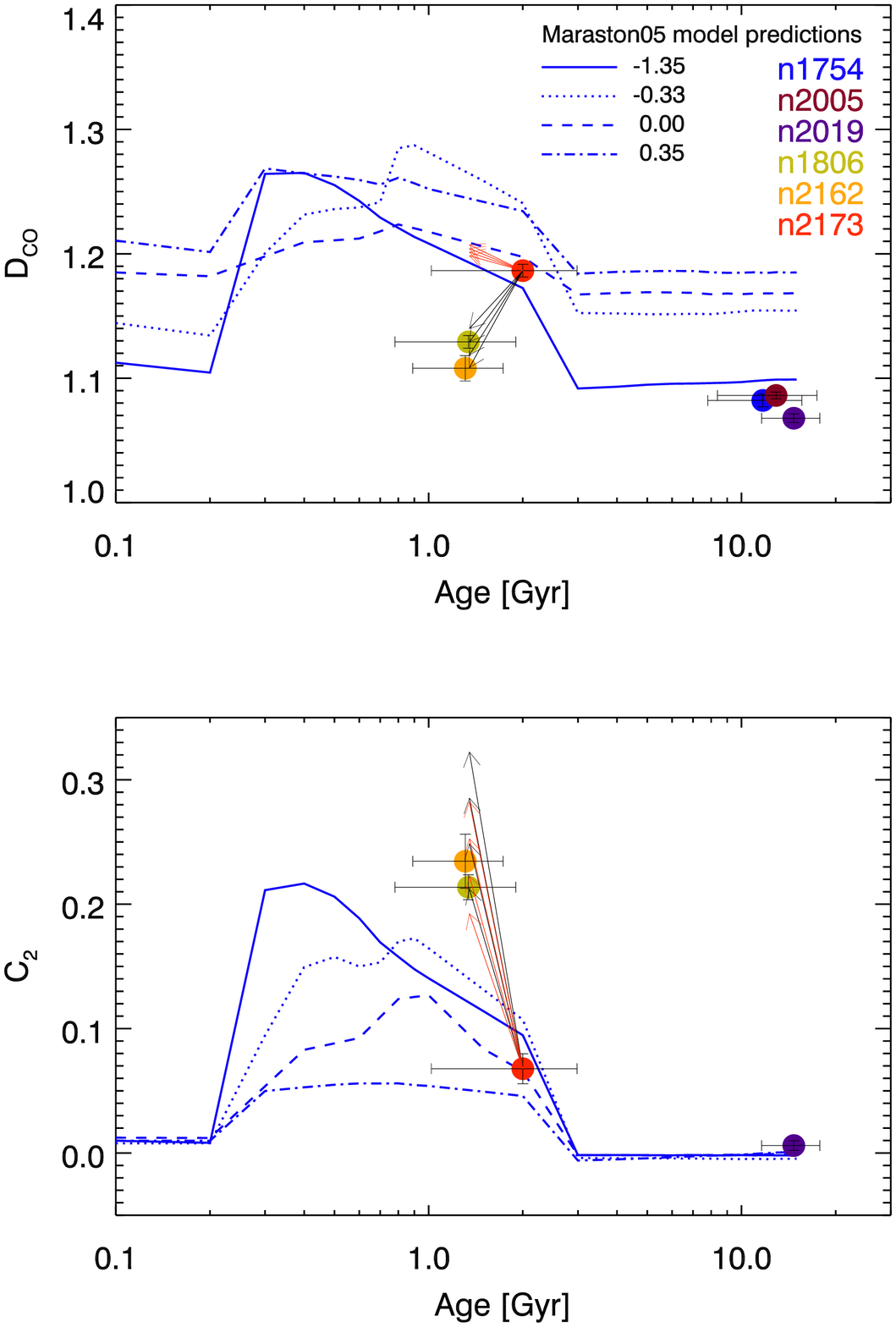}}
\caption{\label{fig:c2_dco_model} Comparison between \ctwoI\/ index
  values and the models of \citet{maraston05}. The ages of the clusters are the mean ones from Table~\ref{tab:lmc_gc} with error bars reflecting the r.m.s.  For guidance, in the
  top panel we reproduce the $K$-band \dco\/ index from \paper1 (their
  Fig.~9). In both panels with black arrows we indicated how the
  spectrum of NGC\,2173 changes when adding different fractions of LMC
  carbon star light to it. The red arrows indicate the changes in the
  same spectrum, but adding different fractions of the averaged Milky
  Way carbon star spectrum of \citet{lancon02}. See text for a
  detailed description.}
\end{figure}

%
%
\begin{center}
\begin{table*}[tdp]
\centering
\caption{\label{tab:bs_observations}Near-IR colours and spectral indices of the additional bright RGB and AGB stars.}
\begin{tabular}{c c c c c c c }
\hline 
\hline 
Name &  $K$ (mag) & $(J-K)$ & \dco & \ctwoI (mag) & Notes & Cluster \\ 
(1) & (2) & (3) & (4) & (5) & (6) & (7) \\
\hline
2MASS\,J04540127-7026341 &  11.40 & 1.13 & 1.208$\pm$0.013  &    0.034$\pm$0.010 &   & NGC\,1754 \\
2MASS\,J04540771-7024398 &  11.86 & 1.26 & 1.223$\pm$0.021  &    0.152$\pm$0.013 &   & '' \\
2MASS\,J04543522-7027503 &  12.26 & 1.07 & 1.143$\pm$0.015  &  -0.011$\pm$0.018 &   & '' \\
2MASS\,J04542864-7026142 &  12.62 & 0.91 & 1.112$\pm$0.018  &    0.052$\pm$0.025 &   & '' \\
2MASS\,J04540935-7024566 &  12.73 & 0.96 & 1.088$\pm$0.021  &  -0.048$\pm$0.020 &   & '' \\
2MASS\,J04541188-7028201 &  13.04 & 0.91 & 1.177$\pm$0.030  &    0.045$\pm$0.009 &   & '' \\
2MASS\,J04540536-7025202 &  13.36 & 0.93 & 1.122$\pm$0.029  &    0.156$\pm$0.012 &   & '' \\
2MASS\,J05302221-6946124 &  10.65 & 1.21 & 1.258$\pm$0.008  &  -0.021$\pm$0.010 &   &  NGC\,2005 \\
2MASS\,J05300708-6945327 &  11.66 & 1.04 & 1.213$\pm$0.012  &  -0.044$\pm$0.018 &   & '' \\
2MASS\,J05295466-6946014 &  11.68 & 1.12 & 1.246$\pm$0.010  &  -0.045$\pm$0.020 &   & '' \\
2MASS\,J05300704-6944031 &  11.80 & 1.03 & 1.251$\pm$0.010  &  -0.039$\pm$0.027 &   & '' \\
2MASS\,J05300360-6944311 &  12.02 & 0.97 & 1.216$\pm$0.013  &    0.063$\pm$0.036 &   & '' \\
2MASS\,J05295822-6944445 &  12.05 & 1.22 & 1.233$\pm$0.016  &  -0.040$\pm$0.041 &   & '' \\
2MASS\,J05320670-7010248 &  10.49 & 1.13 & 1.136$\pm$0.014  &  -0.018$\pm$0.005 &   & NGC\,2019 \\
2MASS\,J05313862-7010093 &  10.56 & 1.39 & 1.196$\pm$0.017  &    0.339$\pm$0.006 & C  & '' \\
2MASS\,J05315232-7010083 &  10.61 & 0.94 & 1.200$\pm$0.016  &  -0.022$\pm$0.005 &   & '' \\
2MASS\,J05321152-7010535 &  10.87 & 1.06 & 1.176$\pm$0.018  &  -0.036$\pm$0.005 &   & '' \\
2MASS\,J05321647-7008272 &  10.95 & 1.13 & 1.183$\pm$0.016  &  -0.008$\pm$0.006 &   & '' \\
2MASS\,J05320418-7008151 &  11.22 & 1.02 & 1.204$\pm$0.024  &  -0.047$\pm$0.007 &   & '' \\
2MASS\,J05021232-6759369 &  10.23 & 1.84 & 1.051$\pm$0.013  &    0.458$\pm$0.007 & +,C  & NGC\,1806 \\ 
2MASS\,J05020536-6800266 &  10.63 & 1.61 & 1.139$\pm$0.015  &    0.500$\pm$0.009 &  +,C & '' \\
2MASS\,J05015896-6759387 &  10.69 & 1.76 & 1.065$\pm$0.013  &    0.481$\pm$0.010 &  +,C & '' \\
2MASS\,J05021623-6759332 &  11.02 & 1.06 & 1.211$\pm$0.012  &  -0.047$\pm$0.007 &  + & '' \\
2MASS\,J05021870-6758552 &  11.32 & 1.11 & 1.267$\pm$0.015  &  -0.071$\pm$0.011 &  + & '' \\
2MASS\,J05021846-6759048 &  11.74 & 1.00 & 1.243$\pm$0.018  &  -0.053$\pm$0.012 &   & '' \\
2MASS\,J05021121-6759295 &  11.97 & 0.96 & 1.197$\pm$0.026  &  -0.045$\pm$0.024 &   & '' \\
2MASS\,J05021137-6758401 &  11.98 & 1.02 & 1.199$\pm$0.023  &  -0.064$\pm$0.016 &   & '' \\
2MASS\,J06002748-6342222 &   9.60 & 1.80 & 1.075$\pm$0.012  &     0.437$\pm$0.005 & +,C  & NGC\,2162 \\
2MASS\,J06003156-6342581 &  11.64 & 1.03 & 1.202$\pm$0.011  &  -0.043$\pm$0.011 & +  & '' \\
2MASS\,J06003316-6342131 &  12.24 & 0.99 & 1.197$\pm$0.016  &  -0.127$\pm$0.017 & +  & '' \\
2MASS\,J06003869-6341393 &  12.26 & 0.99 & 1.184$\pm$0.020  &  -0.038$\pm$0.017 &   & '' \\
2MASS\,J05563892-7258155 &  11.77 & 1.03 & 1.227$\pm$0.012  &  -0.091$\pm$0.013 &  & NGC\,2173 \\
2MASS\,J05575667-7258299 &  12.07 & 0.95 & 1.166$\pm$0.014  &  -0.047$\pm$0.017 & + & '' \\
2MASS\,J05570233-7257449 &  12.13 & 1.04 & 1.159$\pm$0.016  &  -0.008$\pm$0.019 &  & '' \\
2MASS\,J05575784-7257548 &  12.18 & 1.03 & 1.203$\pm$0.016  &  -0.048$\pm$0.019 & + & '' \\
2MASS\,J05563368-7257402 &  12.43 & 1.00 & 1.146$\pm$0.016  &    0.010$\pm$0.027 &  & '' \\
2MASS\,J05581142-7258328 &  12.45 & 1.04 & 1.154$\pm$0.019  &  -0.021$\pm$0.023 & + & '' \\
2MASS\,J05583257-7258499 &  12.48 & 0.92 & 1.136$\pm$0.015  &  -0.070$\pm$0.022 &  & '' \\
2MASS\,J05572334-7256006 &  12.56 & 1.01 & 1.227$\pm$0.019  &  -0.016$\pm$0.027 &  & '' \\
2MASS\,J05565761-7254403 &  12.84 & 1.01 & 1.169$\pm$0.020  &  -0.099$\pm$0.031 &  & '' \\
\hline
\hline
\end{tabular}
\tablefoot{
(1) 2MASS catalogue star name, (2) extinction corrected $K$-band magnitude and (3) $(J-K)$ colour from  the 2MASS Point Source Catalogue \citep{2mass}, (4) $K$-band \dco\/ index value, (5) $H$-band \ctwoI\/ index value, (6) Notes on individual stars: "C" -- a carbon-rich stars, "+" -- the star was used for the integrated spectrum of  the cluster, (7) globular cluster, next to which the star was observed. The \dco\/ index is defined in \citet{esther08} and the \ctwoI\/ index is taken from \citet{maraston05}.}
\end{table*}
\end{center}
%
%

%
%
\begin{table}[tdp]
\caption{\label{tab:lmc_c2} Measurements of \dco\/ and \ctwoI\/ indices for the LMC globular clusters.}
\centering
\begin{tabular}{c c c}
\hline 
\hline 
Name & \dco & \ctwo (mag) \\ 
\hline
NGC\,1806 & 1.129$\pm$0.005 & 0.214$\pm$0.010\\ 
NGC\,2162 & 1.108$\pm$0.010 & 0.235$\pm$0.022 \\ 
NGC\,2173 & 1.186$\pm$0.005 & 0.068$\pm$0.012 \\
NGC\,2019 & 1.068$\pm$0.003 & 0.006$\pm$0.004 \\
NGC\,2005 & 1.086$\pm$0.003 &  --\\
NGC\,1754 & 1.082$\pm$0.005 & --\\
\hline
\hline
\end{tabular}
\tablefoot{
The \dco\/ index is defined in \citet{esther08} and the \ctwoI\/ index is taken from \citet{maraston05}.}
\end{table}

Fig.~\ref{fig:c2_dco_model} shows a comparison between our
measurements of the \ctwoI\/ index for the LMC globular clusters and
the SSP model predictions of \citet{maraston05}. In the top panel, for
guidance, we repeat Fig.~9 from \paper1\/, showing the $K$-band \dco\/
index. This index is sensitive to the metallicity at older ages, but
also shows a dependence on the presence of carbon-rich stars at
intermediate ages. We measured the \dco\/ and \ctwoI\/ indices at the
nominal spectral resolution of the data. Current models have lower spectral resolution than our data. However, due to the broadness and intrinsic strength of these spectral features, the differences in
spectral resolution will not affect our conclusions. We tested
this by broadening our globular cluster spectra to spectral
resolutions ranging from 50 to 400~\kms\/ in steps of 50~\kms\/. The
largest offset we measured was $-0.03$ for both the \dco\/ and \ctwoI\/
indices, which is much smaller than the observed differences between
observations and models.
 
For old stellar populations ($>$3\,Gyr) the models predict an
approximately constant and zero value of the \ctwoI\/ index
(Fig.~\ref{fig:c2_dco_model}, bottom panel). The \ctwoI\/ index
measurement for the old globular cluster NGC\,2019 in this age range
whose metallicity is best compared to the model with [Z/H]~$=-1.35$
(solid, blue line), is consistent with the model predictions. We note
that we were unable to measure the \ctwoI\/ index in the other two old
and metal poor clusters in our sample, due to the very low quality of
their $H$-band spectra.

The group of the three intermediate age and more metal rich clusters
is best compared with the SSP model with [Z/H]~$=-0.33$ (blue, dotted
line). The $C_2$ index measurement for NGC\,2173 (2\,Gyr, red solid symbol) is consistent with the model prediction of increasing $C_2$ index towards younger ages. The youngest   
globular clusters at $\sim$1.3\,Gyr (NGC\,1806 and NGC\,2162) follow this trend, although
their $C_2$ index is significantly larger than the model prediction.
The latter observation supports our findings based on the $K$-band \dco\/
index in \paper1. There, we argued that the reason for the discrepancy
between the CO index predictions of the models and the data (see Fig.~\ref{fig:c2_dco_model}, top panel) is due to the presence of carbon-rich stars which influence the
spectrophotometric properties of stellar populations at intermediate
ages. The empirical calibration of the models is based on spectra of
the Milky Way carbon stars from the library of \citet{lancon02},
combined with photometric calibration based on the LMC globular
clusters, which led to inconsistent results. The presence of LMC
carbon-rich stars in the spectra leads to a decrease of the \dco\
index, but to an increase of the \ctwoI\ index. 
To test this claim, we used the spectrum of NGC 2173, which is the "oldest" intermediate-age globular cluster and therefore the one with the least contribution from carbon-rich stars in our sample \citep{muc06}, as a baseline for a simplistic stellar population modelling test. Increasing the contribution from TP-AGB stars in this cluster would mimic a younger age, such as seen in 
NGC\,1806 and NGC\,2162. In this way we are able to test the effects of different carbon- to oxygen-rich stars ratios in a given stellar population on its spectral features. When we added 40\% to 70\% of the averaged Milky Way stellar spectrum contained in the 3rd bin of \citet{lancon02} to the spectrum of NGC 2173, the resulting \ctwoI\/ and \dco\/ indices increased, as predicted by the models for the younger clusters (red arrows in Fig. 4). When performing the same test, but adding the spectrum of a carbon star in the LMC (in this case the star 2MASS J06002748-6342222 in NGC 2162) to the cluster, the resultant \dco\/ index {\em decreased}\/ (black arrows in the same figure) and fitted the observations of the younger clusters better. Contrary to the \dco\/ index dependence on C-star content, the resultant \ctwoI\/ index increased consistent with the model predictions regardless if the C-star is taken from the Milky Way or the LMC.

The results from the above tests depend on the actual carbon star spectrum that is used to perform the test. Carbon stars exhibit large variations in their properties for a given environment, as we discuss further in Sect.~\ref{sec:discussion} \citep[see also e.g.][]{lancon02}. This caveat illustrates the intrinsic problem of stochastic fluctuations of AGB stars in globular clusters \citep[see e.g. review
by][]{lancon11}. AGB stars are rare in stellar populations, with only
one in a population of 10$^{4}$ stars in total. Nevertheless, a single
star can produce up to 80\% of the stellar population's near-IR
light \citep{maraston05}. Here we see that it can also play a very
important role in the integrated spectral properties of the hosting
stellar population. This will inevitably lead to large or even
significant deviations between the model predictions and observations of
individual clusters. Various authors have made extensive studies
of the influence of such stochastic fluctuations for the integrated
colours of stellar clusters \citep[e.g.][]{piskunov09,popescu10,fouesneau10}. A similar study for the spectral features in integrated spectra of globular clusters would be beneficial for the
interpretation and judgment of our results.

\section{Carbon-rich AGB stars in the LMC and the Milky Way}
\label{sec:discussion}

A possible explanation for the differences between models and
observations, described in the previous section, may be the different
metallicities or carbon and oxygen abundances of the sample stars. In more metal poor stars (i.e. the
LMC stars), the C/O ratio is higher on average. \citet{matsuura02} argue for a
systematically larger C/O ratio in LMC carbon stars compared to
Galactic ones. Models of carbon-rich stars with higher C/O ratio have
weaker CO features, but stronger \ctwoF\/ features
\citep{loidl01,aringer09}. We note that a variation in the molecular
bands of carbon stars during their pulsation cycles is not expected to
be strong \citep{loidl01}. In \paper1\/ we supported this scenario, i.e. the difference in metallicity to be the cause of the discrepancy, by showing that stars in the LMC and the Milky Way with similar $(J-K)$ colours can have substantially different \dco\/ index values, as \citet{cohen81} and \citet{frogel90} have shown earlier as well.

\begin{figure*}
\resizebox{\hsize}{!}{\includegraphics[angle=0]{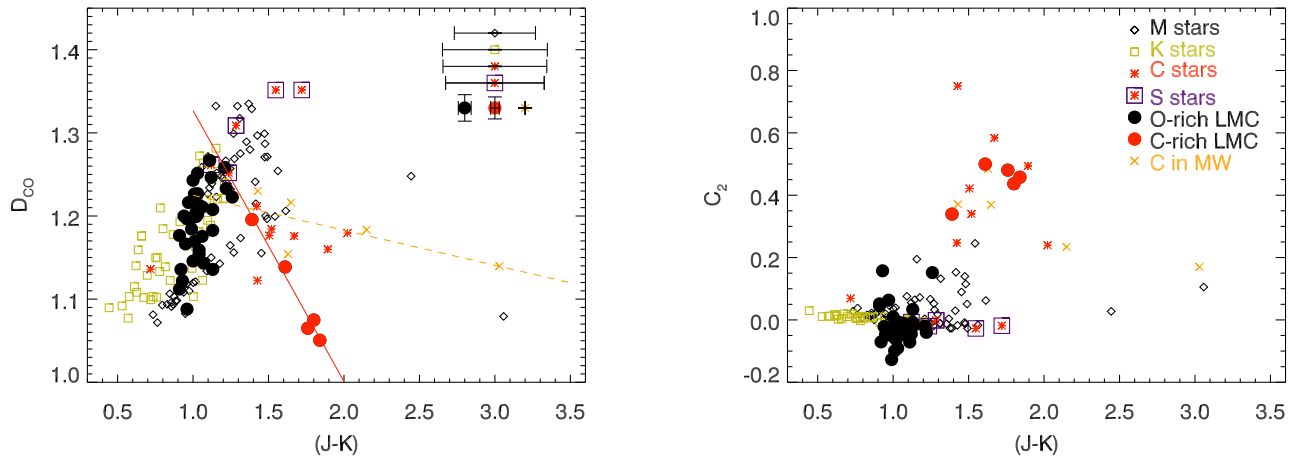}}
\caption{\label{fig:dco_c2_jk_stars} \dco\/ {\it (left)} and \ctwoI\/
  {\it(right panel)} indices versus $(J-K)$ colour for stars in the
  Milky Way and the LMC. The oxygen- and carbon-rich LMC stars (filled
  black and red dots, respectively) are from our own SINFONI
  observations, the K (green open squares), M (black diamonds), C (red
  asterisks), and S (red asterisks with square around) stars are taken
  from the SpeX library of \citet{rayner09}. The C-rich spectra in the
  Milky Way (orange crosses) are taken from the library of
  \citet{lancon02}. Typical error bars are shown in the upper right
  corner of the left panel. In the left panel, the solid line denotes
  the relation for LMC carbon-rich stars from \citet[][Eq.~3]{az10a},
  the dashed line (Eq.~5 in the same paper) represents the relation
  for the Milky Way spectra of \citet{lancon02}.}
\end{figure*}

Based on the limited stellar samples discussed there, we speculate that the
LMC carbon-rich stars have weaker \dco\/ indices when compared to
carbon stars in the Milky Way. In Fig.~\ref{fig:dco_c2_jk_stars} (left
panel) we repeat the diagram and add stars from the SpeX library
\citep{rayner09}. This library contains spectra of mostly near-solar
metallicity K and M stars with luminosity classes between I and V,
carbon-rich stars and S-stars (for which C/O$\simeq$1). For half of the
carbon-rich stars (four) we could find iron abundance estimates in the
PASTEL database \citep{soubiran10}, ranging from $-0.3$ to 0.2
dex. Their mean iron abundance is slightly sub-solar, which is on
average more metal-rich than the sample of LMC carbon stars with
[Fe/H]~$\simeq-0.4$. The carbon-rich stars from the SpeX sample are
located between the two relations, found in \paper1, for LMC (solid
line) and Milky Way (dashed line) carbon stars from the library of
\citet{lancon02}, thus smearing any clear relations. Moreover, if instead of the averaged Milky Way carbon star spectra, the individual stars of \citet{lanconW00} were plotted, then the dispersion would have been even more prominent. However, we do not observe \dco\/ in any of the Milky Way stars as weak as in three
of the LMC stars. Surprisingly, on the $(J-K)$ -- \ctwoI\/ plot
(Fig.~\ref{fig:dco_c2_jk_stars}, right panel) there is no obvious
difference between the stars in the two galaxies. Assuming the lower
metallicity of the LMC and the higher C/O ratio, as argued by
\citet{matsuura02}, the \ctwoI\/ index is expected to be stronger in
the LMC than in the Milky Way carbon stars. Instead we find the
opposite: that two of the Galactic stars have a stronger \ctwoI\/ index
than the LMC stars. This could be due to the C/O ratio changing at every dredge-up episode, thus leading to a dispersion in the \ctwoI\/  index larger than the metallicity effect. Based on the three stellar samples explored here, we cannot confirm that there is a marked difference between the Milky
Way and LMC carbon stars in terms of the C/O ratio.  To reach
conclusive results about the dependence of the \ctwoI\/ index on the
C/O ratio, different element abundances, and the metallicity of the stars, detailed abundance
measurements for a larger sample of LMC and Milky Way giant stars are
necessary.

\begin{figure}
\resizebox{\hsize}{!}{\includegraphics[angle=0]{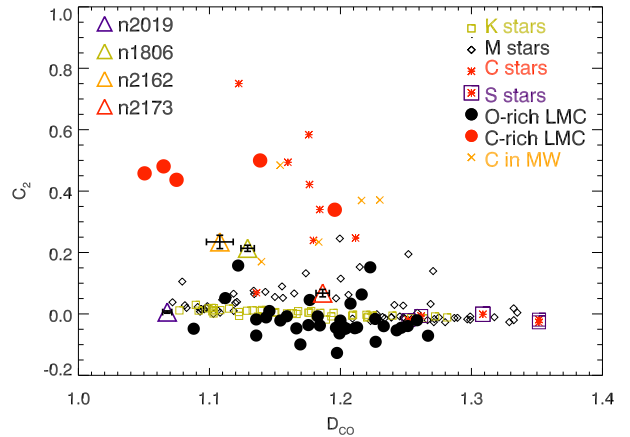}}
\caption{\label{fig:dco_c2_stars} \dco\/ vs. \ctwoI\/ index for stellar
  and globular cluster spectra. Origin of spectra, symbols and colours
  as in Fig.~\ref{fig:dco_c2_jk_stars}.  Error bars are about the size
  or smaller than the symbols, thus they are not explicitly shown.
  Coloured triangles show the measurements for the integrated spectra
  of LMC globular clusters.}
\end{figure}

Fig.~\ref{fig:dco_c2_stars} shows the behaviour of the $K$-band CO
versus the $H$-band \ctwoF\/ features in stellar spectra.  The complete
sample of stars displays a large spread of \dco\ index values, while
the \ctwoI\ index is significantly stronger than zero only in
carbon-rich stars. There are also a number of M-type stars (M6-9III),
whose \ctwoI\/ index is larger than zero but is still systematically
smaller than the value of \ctwoI\/ in C-type stars. In these stars
there is significant $H_{2}O$ absorption in the range $1.7 -
2.05\,\mu$m \citep[e.g.][]{matsuura99,rayner09} that leads to a steep
decrease of the continuum at the location where the \ctwoF\/ feature is
located. Thus, when measuring the \ctwoI\/ index in these M giants we
obtained a non-zero index value, while in reality there is no \ctwoF\/
absorption.
    
Based on this figure, there is no clear separation between LMC and
Milky Way carbon-stars, although the LMC stars have on average lower
\dco\/ indices, which is consistent with their lower
metallicity. However, the same LMC stars do not show a stronger
\ctwoI\/ index, which would be expected as discussed earlier in the
text.

There is one C-type star from the catalogue of \citet{rayner09} that
shows \dco\/ and \ctwoI\/ indices and a $(J-K)$ colour consistent with
K and M-type stars (see Fig.~\ref{fig:dco_c2_jk_stars}). This is
HD\,76846, classified as C-R2 \citep{keenan00}.  As \citet{tanaka07}
have shown, the \ctwoF\/ feature is intrinsically weak in early hotter
C-R stars.

In Fig.~\ref{fig:dco_c2_stars}, for comparison, we show the measured
\dco\/ and \ctwoI\/ indices from the integrated LMC globular cluster
spectra (coloured triangles). The two clusters that are dominated by
carbon stars, NGC\,1806 and NGC\,2162, follow the stellar trends of
higher \ctwoI\/ index values. Such a near-IR spectroscopic index diagnostic plot can
be very helpful in revealing the age of a given integrated stellar
population, being a stellar cluster or a galaxy. As shown above, the
\ctwoF\/ feature is present and its strength is significant only in
carbon stars and stellar populations harbouring such.  \citet{lancon99} have already proposed a similar \ctwo\/ index, that can be used to detect the presence of intermediate age stellar populations.  Even when
smeared out by the line-of-sight velocity distribution of stars in
galaxies of the order of 400\,\kms\/, it will remain identifiable and
measurable. By observing an increased strength of the \ctwoI\/ index,
one can conclude that the object is in the carbon-star phase and thus
date it to an intermediate age of 1-2\,Gyr. 

%
%
%
%

%
%

\section{Concluding remarks}
\label{sec:conclusions}
In this paper we concluded our pilot project of providing an empirical
near-IR library of integrated spectra of globular clusters in the LMC
by adding $J$- and $H$-band spectra to the $K$-band ones from
\citet[][\paper1]{az10a}. We provided full near-IR spectral coverage
(as observable from the ground) for four globular clusters: the old and
metal poor cluster NGC\,2019, and the intermediate age and
approximately half-solar metallicity clusters NGC\,1806, NGC\,2162,
and NGC\,2173. For the old and metal poor clusters NGC\,1754 and
NGC\,2005 the $H$-band part is missing due to bad observing conditions
and unacceptable sky emission subtraction residuals. Using our sample
of globular clusters we tested predictions of current evolutionary
population synthesis models in the near-IR and discussed some of the
problems that arise from this comparison. Although still influenced by
small number statistics, our spectra show that the near-IR spectral
domain is a useful source of information when applied to the study of
integrated stellar populations.

One of the main goals of our project was to provide a small empirical
library of near-IR spectra of globular clusters to serve as a test
bench for current and future stellar population models. In
Fig.~\ref{fig:lmc_sed} we presented a comparison between the overall
near-IR spectral energy distribution of sample clusters and the
stellar population models of \citet{maraston05}. The models agree
remarkably well with the data in terms of the spectral shape and the
most prominent absorption features. Especially, the predicted distinct
signatures of the TP-AGB phase in intermediate age (1-2\,Gyr) stellar
populations are nicely reproduced.

However, the agreement is not so good when exploring individual
spectral features in more detail, where more subtle differences in
absorption strength can be quantified. Here, as well as in \paper1\/,
we concentrated our studies to the strongest absorption features in
the near-IR, namely \ctwoF\/ at 1.77\,$\mu$m and \co\/ at 2.29\,$\mu$m,
due to the limited spectral resolution of the current models. We
described their behaviour using two indices: \ctwoI\/ and \dco\/,
respectively. The first one reflects the presence of carbon-rich AGB
stars in the stellar population, while the second is typical for both
carbon- and oxygen- rich AGB and RGB stars. We refer the reader to Sect.~\ref{sec:c2}, where we discussed in detail the observed differences between our observations and the EPS models of \citet{maraston05}. In Sect.~\ref{sec:discussion} we explored some possible reasons for these, based on stellar libraries with Milky Way and LMC carbon-rich stars (see Fig.~\ref{fig:dco_c2_jk_stars}). These stars can contribute up to 80\% to the total $K$-band light of a globular cluster stellar population, thus their proper inclusion in models is crucial. Our results were inconclusive about the proposed difference in the C/O ratio between stars in the Galaxy and the LMC. To confirm or reject this, accurate abundance measurements of iron, oxygen and carbon elements are necessary. Therefore, there is a clear need for larger spectral libraries of carbon-rich stars with a range of metallicities before we are able to reproduce their proper contribution to the stellar population models.

The occurrence of a strong \ctwoI\/ index in the near-IR spectra of
globular clusters and hence in galaxies is an indication that they are
in the carbon star phase, which dates the hosting stellar population
to $1-2$\,Gyr, as previously suggested by \citet{lancon99}. In contrast, the \dco\/ index is much less
straightforward to interpret, since the magnitude of this index is
driven by a complex combination of metallicity, surface gravity
(luminosity), and effective temperature, as illustrated by
Fig.~\ref{fig:dco_c2_jk_stars}. In Fig.~\ref{fig:dco_c2_stars} we
propose a diagnostic plot, based only on these two near-IR
indices. Both of them are sufficiently broad and strong, so they can
be easily measured in galaxy spectra, even when smeared out by the
line-of-sight velocity distribution of the stars. For galaxies up to a
redshift of 0.007, e.g. Virgo and Fornax clusters, the \ctwoF\/ feature
is in the rest-frame $H$-band. For redshifts between 0.007 and 0.13
the feature is hidden by the atmospheric cutoff, but it remains
accessible from space.

During the next decades new facilities, such as the JWST and the
E-ELT, will widen the discovery space in extragalactic research. Their
enhanced sensitivity will be in the infrared portion of the
spectrum. However, the stellar population analysis methods for the
integrated light are not yet fully developed in the near-IR wavelength
regime.  The methods discussed here offer a step towards a better
understanding of galaxy evolution and observations are already
feasible with current facilities.  For example, current adaptive
optics systems, working mostly in the near-IR, offer a possibility to
resolve the innermost parts of nearby galaxies.  New information about
the stellar population properties at these spatial scales can give us
important clues about how galaxies have formed and evolved.

\acknowledgements{We are grateful to the many ESO astronomers who
  obtained the data presented in this paper in service mode operations
  at La Silla Paranal Observatory. We thank Livia Origlia for useful
  discussions on near-IR spectral synthesis and Claudia Maraston for her comments. ML would like to thank ESO's Visitors Program as well as the staff at the Astronomical
  Observatory of the University of Sofia for their hospitality, where
  major parts of this research have been carried out. Finally, we thank the referee for her/his helpful suggestions which certainly made this paper more complete. This paper is dedicated to Ralitsa Mitkova, a tiny star that evolved together with this manuscript and ever since provides an endless source of inspiration.}

\bibliographystyle{aa}
\bibliography{/Users/mlyubeno/Dropbox/sci/biblio/papers}

\end{document}